\documentclass[a4paper,10pt]{article}

\title{Decoherence of quantum states and its suppression in
ensemble large-scale solid state NMR quantum computers}
\author{\large{A.\ A.\ Kokin}}
\date{}

\begin{document}

\maketitle

\thanks{\large Institute of Physics and Technology of RAS, 34, Nakhimovskii pr., 117218 Moscow,
Russia}

\begin{abstract} \normalsize
It is discussed the decoherence problems in ensemble large-scale
solid state NMR quantum computer based on the array of
$^{31}\mathrm{P}$ donor atoms having nuclear spin $I = 1/2$. It is
considered here, as main mechanisms of decoherence for low
temperature ($< 0.1\,\mathrm{K}$), the adiabatic processes of random
modulation of qubit resonance frequency determined by secular part
of nuclear spin interaction with electron spin of the basic atoms,
with impurity paramagnetic atoms and also with nuclear spins of
impurity diamagnetic atoms. It was made estimations of allowed
concentrations of magnetic impurities and of spin temperature
whereby the required decoherence suppression is obtained. It is
discussed the random phase error suppression in the ensemble
quantum register basic states.
\end{abstract}

keywords: decoherence, ensemble, large-scale, nuclear magnetic
resonance, qubit, quantum computer.

\section*{Introduction}

There are five basic criteria for realization of a large-scale
quantum computer, which can outperform all traditional classical
computers \cite{1}. Let us take a brief look at these criteria in
respect to the large-scale NMR quantum computers.\par

1. For any physical system, which presents large-scale quantum
register, it is required of a sufficiently large qubit number ($L
> 10^{3})$ and each qubit should be separately identifiable.\par

One such example of this register is solid-state homonuclear
system, in which nuclear spin containing identical atoms are
housed at regular intervals in a natural or an artificial
solid-state structure.\par

2. There is a need to provide the conditions for preparation of
initial and basic quantum register state. For a many-qubit NMR
quantum computer the quantum register state initializing can be
obtained by going to extra-low nuclear spin temperature ($<
1\,\mathrm{mK}$ at fields of order of several Tesla).\par

3. The decoherence time of qubit states $T_{\mathrm{d}}$ should be
at least up to $10^{4}$ time longer than the ''clock time''. It
has value of order of several seconds for NMR quantum computers.
The decoherence suppression is one of the important problems in
realization of a large-scale quantum computers.\par

4. There is a need to provide during a decoherence time the
implementing of a set of quantum logic operations determined by a
logic unitary transformation. This set should contain certain set
of the one-qubit and two-qubits operations are shielded from
random errors. The electromagnetic pulses that control the quantum
operation should be performed with an accuracy of better than
$10^{-4}$--$10^{-5}$.\par

5. There is a need to provide the accurate and sensitive read-out
projective measurements of individual qubit states. This is
another of the important and the most hard problems.\par

One of the pioneering schemes of large-scale NMR quantum computer with
individual access to qubits was proposed by B.Kane \cite{2,3}.\par

There are four peculiar difficulties in implementing of such
quantum computer:\par

1. First of all, signal from the spin of an individual atom is
very small and it is required of a highly sensitive single-spin
measurements.\par

2. For initialization of nuclear spin states it is required to use
very low nuclear spin temperature ($\sim \,\mathrm{mK}$).\par

3. It is required to use the regular donors and gates arrangement
with high precision in nanometer scale.\par

4. It is necessary to suppress the decoherence mechanism defined
by thermal fluctuations of gate voltage (this mechanism will be
denoted as external).\par

As an alternative, we have proposed the variant of an ensemble
silicon-based quantum computer \cite{4,5,6}. One would expect that
with the ensemble approach when many independent ''molecules'' of
Kane's type, like in the bulk-ensemble liquid quantum computers,
work simultaneously, the qubit state measurements seem to be
greatly simplified. Furthermore, it seems to be solved certain
problems associated with possible noncontrolled continuous random
phases of individual qubits.\par

This article discusses only the decoherence mechanisms of one
qubit quantum states in ensemble quantum register based on the
regular array of $^{31}\mathrm{P}$ atoms having nuclear spin $I =
1/2$. It is considered, as main mechanisms of decoherence for the
very low temperature, the adiabatic processes of random modulation
of qubit resonance frequency without spin flips produced by
fluctuating local magnetic fields which are determined by secular
parts of interactions of nuclear spins with electron spin of the
basic atoms, with impurity paramagnetic atoms and also with
nuclear spins of impurity diamagnetic atoms. We have named this
mechanisms as internal. It was made estimations of allowed
concentrations of magnetic impurities that are necessary for
decoherence suppression. It is discussed also the random phase
error suppression in the initializing process for ensemble quantum
register states.\par

\section{Mechanisms of qubit state decoherence in NMR
solid-state quantum registers}

The full electron spin polarization will be practically achieved
at
\begin{eqnarray}
\gamma _{\mathrm{S}}\hbar B/kT \gg 1 ,\label{1}
\end{eqnarray}
where $\gamma _{\mathrm{S}} = 176\,\mathrm{rad}\,
\mathrm{GHz}/\mathrm{T}$ is giromagnetic ratio for electron spin,
$\hbar = 1.05\cdot 10^{-34}\,\mathrm{J}\,
\mathrm{s}/\mathrm{rad}$, $B$ is induction of external magnetic
field, $T$ is temperature of environment (of electron spins). Thus
for $B = 2\,\mathrm{T}$, $T = 0.1\,\mathrm{K}$ ($B/T = 20\,\mathrm{T}/\mathrm{K})$ we
obtain $\gamma _{\mathrm{S}}\hbar B/kT = 27$.\par

The electron and nuclear longitudinal relaxation processes, that
are followed by spin flips in four energy level system of
phosphorus atoms doped in silicon, have been previously
investigated experimentally in \cite{7}. The electron longitudinal
relaxation times at low temperatures were found to be extremely
long (hours) and independent of phosphorus concentration below
$\sim 10^{16}\,\mathrm{cm}^{-3}$. The nuclear longitudinal
relaxation time $T_{\parallel} $ were found to be equal to 10
hours \cite{7}.\par

The extremely long longitudinal relaxation times of the electron
and nuclear spins imply the possibility to produce a
long-lived nuclear nonequilibrium initialized state for lattice
temperatures (say, for $T = 0.1\,\mathrm{K})$ when only the
electron spins are polarized. The required initializing of nuclear
quantum states (near-full nuclear polarizations) can be obtained
by a short duration deep cooling to $T_{\mathrm{I}} \leq
1\,\mathrm{mK}$ only for nuclear spin system without deep cooling of
the whole sample. There is the possibility to achieve it by
indirect cooling of nuclear spin system using dynamic nuclear spin
polarization techniques (such as the known Abragam's solid-state
effect) \cite{8}.\par

The relaxation of nonequilibrium state of the nuclear spin system
represented by the product of independent (nonentangled) one-qubit
states, owing to the interaction with isotropic environment, shows
two processes. One is a slow establishment of equilibrium state
associated with dissipation of energy. For it the diagonal
elements of density matrix decay with characteristic longitudinal
(spin-lattice) relaxation time $T_{\parallel} $. The decay of
non-diagonal matrix elements called decoherence of quantum states
is characterized by a decoherence time $T_{\mathrm{d}}$ or
transverse (spin-spin) relaxation time $T_{\perp }$. The
longitudinal relaxation times $T_{\parallel} $ in the case of
nuclear spin of $^{31}\mathrm{P}$ atoms as qubits is defined
mainly by thermal modulation of hyperfine interaction accompanied
by spin flips. It is usual that for solids $T_{\perp } \ll
T_{\parallel} $.\par

We will not consider here the external decoherence process due to
gate voltage noise. It was made in \cite{2,3,9}.\par

The internal adiabatic decoherence mechanisms due to a random
modulation of qubit resonance frequency, produced by local
fluctuating magnetic fields without spin flips, seem to be the
leading.\par

\section{Semiclassical model of adiabatic decoherence of
one-qubit state}

We will consider a long-lived non-equilibrium qubit state when
diagonal elements of density matrix may be treated as a
constant.\par

The random modulation of resonance frequency $\Delta \omega (t)$
that causes the dephasing of a qubit state are determined by the
random phase shifts
\begin{eqnarray}
\varphi (t) = \int_{0}^{t}\Delta \omega (t)dt.\label{2}
\end{eqnarray}
The one-qubit density matrix of pure state in rotating frame
with non perturbed resonance circular frequency will be
\begin{eqnarray}
\rho (t) = 1/2\left[
\begin{tabular}{c c }
 $1 + P_{\mathrm{z}}$ & $P_{-}\exp (i\varphi (t))$ \\
 $P_{+}\exp (-i\varphi (t))$ & $1 - P_{\mathrm{z}}$ \\
\end{tabular}
\right],\label{3}
\end{eqnarray}
where $P_{\pm } = P_{\mathrm{x}} \pm iP_{\mathrm{y}}$,
$P_{\mathrm{x}}, P_{\mathrm{y}}, P_{\mathrm{z}}$ are Bloch vector
components of length $P = \sqrt{P_{\mathrm{x}}^{2} +
P_{\mathrm{y}}^{2} + P_{\mathrm{z}}^{2}} = 1$.\par

By treating the resonance frequency modulation as Gaussian random
process after averaging (\ref{3}) over phase distribution with
$\left<\varphi (t)\right> = 0$ we obtain
\begin{eqnarray}
\left<\rho (t)\right> = 1/2\left[
\begin{tabular}{c c }
 $1 + P_{\mathrm{z}}$ & $P_{-}\exp (-\Gamma (t))$ \\
 $P_{+}\exp (-\Gamma (t))$ & $1 - P_{\mathrm{z}}$ \\
\end{tabular}
\right],\label{4}
\end{eqnarray}
where
\begin{eqnarray}
\Gamma (t) = 1/2\cdot \left<(\int_{0}^{t}\Delta \omega
(t)\mathrm{d}t)^{2}\right> = \int_{0}^{t}(t-\tau )\left<\Delta
\omega (\tau )\Delta \omega (0)\right>d\tau ,\label{5}
\end{eqnarray}
$f(t) = \left<\Delta \omega (t)\Delta \omega (0)\right>$ is the
frequency correlation function of a random process, which is
characterized by variance $\left<\Delta \omega (0)^{2}\right>$ and
correlation time $\tau _{\mathrm{C}}$ such that for $t > \tau
_{\mathrm{C}}$ $\left<\Delta \omega (t)\Delta \omega (0)\right>
\Rightarrow 0$. For $\Gamma (t) > 0$ the averaged density matrix
presents a mixed quantum state with two non-zero eigen states
\begin{eqnarray}
1/2\cdot \left( 1 \pm \sqrt{1 - (P_{\mathrm{x}}^{2} +
P_{\mathrm{y}}^{2})(1-\exp (-2\Gamma (t))}\right)\label{6}
\end{eqnarray}
and the populations of states $p_{\pm } = 1/2(1 \pm
P_{\mathrm{z}}(0))$ at $\Gamma (t) \Rightarrow \infty $.\par

Thus, the adiabatic decoherence problem is reduced to the
determination of the function $\Gamma (t)$ or the correlation
function of random frequency modulation.\par

In the case of an ensemble quantum register there is a need to
average the one-qubit density matrix and correlation function over
ensemble of independed equivalent spins-qubits.\par

\section{The nuclear spin states decoherence due to hyperfine
interaction of nuclear and electron spins}

In this case the modulation of nuclear spin resonance frequency
$\Delta \omega (t)$ is determined by the secular part of hyperfine interaction:
\begin{eqnarray}
\Delta \omega (t) = A(t) S_{\mathrm{z}}(t) -
A_{0}\left<S_{\mathrm{z}}\right>.\label{7}
\end{eqnarray}
The most studied exactly solvable one-boson models of adiabatic
decoherence that is not followed by spin flips are not adequate
\cite{10}. The decoherence that is described by the direct
one-phonon process with spin flips is non-adiabatic slow
relaxation process. It is characterized by a rate comparable to
1/$T_{\parallel} $. To describe the dephasing or pure adiabatic
decoherence we will take in to account one of the Raman two-phonon
processes, which describes the hyperfine interaction constant
modulation $A(t)$ due to the phonon scattering without of spin
flip. In interaction representation we will have
\begin{eqnarray}
A(t) = A_{0}[1 + {\sum_{\mathrm{m}\neq
\mathrm{l}}}\mathrm{g}_{\mathrm{ml}}b_{\mathrm{m}}^{+}b_{\mathrm{l}}\exp
(i(\omega _{\mathrm{m}}-\omega _{\mathrm{l}})t)
+\,\mathrm{g}^{*}_{\mathrm{ml}}b_{\mathrm{l}}^{+}b_{\mathrm{m}}\exp
(-i(\omega _{\mathrm{m}}-\omega _{\mathrm{l}})t)],\label{8}
\end{eqnarray}
where $A_{0} = 725\,\mathrm{rad}\, \mathrm{MHz}$ is hyperfine
interaction constant for atoms $^{31}\mathrm{P}$ at $T = 0$,
$b_{\mathrm{l}}^{+}$ and $b_{\mathrm{m}}$ are the emission and
absorption of phonon operators with frequencies $\omega
_{\mathrm{l}}$ and $\omega _{\mathrm{m}}$, $g_{\mathrm{ml}} \sim
\xi (\hbar /2MvN)(\omega _{\mathrm{m}}\omega _{\mathrm{l}})^{1/2}$
is the coupling constants of hyperfine parameter with acoustic
phonons, dimensionless parameter $\xi $ in the worse case is $\sim
1$, $v$ is the velocity of sound, $N$ is the total number of atoms
in crystal (for simplicity we assume the simple cubic lattice),
$M$ is the atom mass.\par

From (\ref{7}) we obtain the sum of two independent terms, which
correspond to the two mechanisms of adiabatic decoherence due to
hyperfine interaction:
\begin{eqnarray}
\Delta \omega (t) &=& \Delta \omega _{\mathrm{S}}(t) + \Delta
\omega _{\mathrm{b}}(t) = A_{0}(S_{\mathrm{z}}(t) -
\left<S_{\mathrm{z}}\right>) +  \nonumber
\\
&& + 1/2 \cdot A_{0} {\sum_{\mathrm{m}\neq
\mathrm{l}}}\mathrm{g}_{\mathrm{ml}}b_{\mathrm{l}}b_{\mathrm{m}}^{+}\exp
(i\omega _{\mathrm{ml}}t)
+\,\mathrm{g}^{*}_{\mathrm{ml}}b_{\mathrm{l}}^{+}b_{\mathrm{m}}\exp
(-i\omega _{\mathrm{ml}}t),\ \ \  \omega _{\mathrm{ml}} = \omega
_{\mathrm{m}} -\omega _{\mathrm{l}}. \label{9}
\end{eqnarray}
At first, let us consider the first term in (\ref{9}). The
correlation function is determined by the fluctuations of electron
spin polarization and depends on electron resonance frequency
$\omega _{\mathrm{S}}$, longitudinal $\tau _{1}$ (hours) and
transverse $\tau _{2}$ relaxation times. In adiabatic case $\omega
_{\mathrm{S}} = \gamma _{\mathrm{S}}B > 1/\tau _{1} \gg 1/\tau
_{2}$ and we will obtain:
\begin{eqnarray}
\left<\Delta _{\mathrm{S}}\omega (t)\Delta \omega
_{\mathrm{S}}(0)\right> = \left<\Delta \omega
_{\mathrm{S}}^{2}\right>\cdot \exp (-t/\tau _{1}),\label{10}
\end{eqnarray}
where
\begin{eqnarray}
\left<\Delta \omega _{\mathrm{S}}^{2}\right> =
A_{0}^{2}(\left<S_{\mathrm{z}}^{2}\right> -
\left<S_{\mathrm{z}}\right>^{2} = A_{0}^{2}(1 - \tanh^{2}(\gamma
_{\mathrm{S}}\hbar B/kT))/4.\label{11}
\end{eqnarray}
Now
\begin{eqnarray}
\Gamma (t) = \left<\Delta \omega _{\mathrm{S}}^{2}\right>\tau
_{1}^{2}(t/\tau _{1} - 1 +\,\exp (-t/\tau _{1})).\label{12}
\end{eqnarray}
For $\tau _{1} \approx 10^{4}\,\mathrm{s}$ and $t \sim
T_{\mathrm{d}} = 1\,\mathrm{s}, 1 \ll \left<\Delta \omega
_{\mathrm{S}}^{2}\right>\tau _{1}^{2} < (\tau
_{1}/T_{\mathrm{d}})^{2}$ we have the non-Markovian random process
(slow damping fluctuations). In this case
\begin{eqnarray}
\Gamma (t) = \left<\Delta \omega _{\mathrm{S}}^{2}\right>t^{2}/2,\
\ \ \ \ \ \ \ \left<\Delta \omega _{\mathrm{S}}^{2}\right> \approx
1/T_{\mathrm{d}}^{2}.\label{13}
\end{eqnarray}
Let us write the requirement for the decoherence time for
$\gamma _{\mathrm{S}}\hbar B/kT \gg 1$ in the form
\begin{eqnarray}
1/T_{\mathrm{d}}^{2} \approx A_{0}^{2}(1 - \tanh^{2}(\gamma
_{\mathrm{S}}\hbar B/kT))/4 \approx A_{0}^{2}\exp (-\gamma
_{\mathrm{S}}\hbar B/kT) < 1\,\mathrm{s}^{-2},\label{14}
\end{eqnarray}
from which we find that the required decoherence suppression will be
achieved at sufficiently large ratio $B/T > 30\,\mathrm{T}/\mathrm{K}$.
For $B/T = 20\,\mathrm{T}/\mathrm{K}$ we have $T_{\mathrm{d}} \sim
10^{-3}\,\mathrm{s}$.\par

For decoherence description due to the second term in (\ref{9}) it
is convenient to use directly the function $\Gamma (t)$. With
\begin{eqnarray}
&& \int_{0}^{t}\Delta \omega _{\mathrm{b}}(t)\mathrm{d}t =
1/2 \cdot A_{0}{\sum_{\mathrm{m},\mathrm{l}}}(\xi
_{\mathrm{ml}}(t)b_{\mathrm{m}}^{+}b_{\mathrm{l}} + \xi
^{*}_{\mathrm{ml}}(t)b_{\mathrm{l}}^{+}b_{\mathrm{m}}), \nonumber
\\
&& \xi _{\mathrm{ml}}(t) = g_{\mathrm{ml}}i(1 -\,\exp (i\omega
_{\mathrm{ml}}t)/\omega _{\mathrm{ml}},\ \ \ \
\left<b^{+}_{\mathrm{m}}b_{\mathrm{m}}\right> = n(\omega
_{\mathrm{m}},T) = (\exp (\omega _{\mathrm{m}}/T - 1)^{-1} \nonumber
\end{eqnarray}
we will
obtain
\begin{eqnarray}
\Gamma (t)  &=& 1/4\cdot \left<\left(\int_{0}^{t}\Delta \omega
_{\mathrm{b}}(t)\mathrm{d}t\right)^{2}\right> = \nonumber
\\ &=&
1/4 \cdot A_{0}^{2}\sum_{\mathrm{m}\neq \mathrm{l}}(\hbar
/2Mv^{2}N)^{2}\cdot (\omega _{\mathrm{m}}\omega
_{\mathrm{l}})(2n(\omega _{\mathrm{m}},T)n(\omega
_{\mathrm{l}},T)+ n(\omega _{\mathrm{m}},T)+ n(\omega
_{\mathrm{l}},T))\cdot \frac{\sin ^{2}(\omega
_{\mathrm{ml}}t)}{\omega ^{2}_{\mathrm{ml}}}.\label{15}
\end{eqnarray}
Let us go now from sums to integrals:
\begin{eqnarray}
{\sum_{\mathrm{m}}}\ldots \Rightarrow \frac{9N}{\Omega ^{3}}
\int_{0}^{\Omega }\ldots \omega ^{2}_{\mathrm{m}}d\omega
_{\mathrm{m}},\label{16}
\end{eqnarray}
where $\Omega = v(6\pi )^{1/2}/a = k\Theta /\hbar$ is Debye
frequency, $a$ is the lattice constant.\par

Taking into account that for $\omega
_{\mathrm{ml}} \Rightarrow 0$ (near-elastic scattering)
\begin{eqnarray}
\frac{\sin ^{2}(\omega _{\mathrm{ml}}t)}{\omega
^{2}_{\mathrm{ml}}} \approx \pi |t|\delta (\omega
_{\mathrm{ml}}),\label{17}
\end{eqnarray}
we now obtain
\begin{eqnarray}
\Gamma (t) = \frac{81\pi }{8} \mathrm{A}_{0}^{2} (\hbar
/Mv^{2})^{2}(T/\Theta )^{7}(k\Theta /\hbar )\int_{0}^{\Theta
/T}\frac{x^{6}\exp x}{(\exp x - 1)^{2}} dx |t| =
|t|/T_{\mathrm{d}}.\label{18}
\end{eqnarray}
For silicon: $\Theta = 625\,\mathrm{K}$, $a = 5.4\cdot
10^{-8}\,\mathrm{cm}$, $v \approx 5\cdot
10^{5}\,\mathrm{cm}/\mathrm{s}$, $M = 0.46\cdot
10^{-29}\mathrm{Js}/\mathrm{cm}^{2}$.\par

At low temperature $T/\Theta < 10^{-3}$ for $T_{\mathrm{d}}$
estimation we obtain:
\begin{eqnarray}
1/T_{\mathrm{d}} \approx \frac{81\pi 6!}{8} \mathrm{A}_{0}^{2}
(\hbar /Mv^{2})^{2}(k\Theta /\hbar )(T/\Theta )^{7} \sim 3/4 \cdot
10^{4} (T/\Theta )^{7} \mathrm{s}^{-1}  \ll 1\,\mathrm{s}^{-1}.\label{19}
\end{eqnarray}
Thus, we see, that at low temperatures the phonon mechanism have
an insignificant effect as compared with the mechanism of electron
spin fluctuations.\par

\section{The nuclear spin states decoherence due to interaction
with magnetic impurity atoms.}

An alternative reason for the modulation of individual nuclear
spin resonance frequency is the secular part of their
dipole-dipole interaction with random distributed paramagnetic
center (impurity atom and defect) in the substrate, which for the
magnetic impurity atoms in S-state is of the form:
\begin{eqnarray}
\Delta \omega (t) = \gamma _{\mathrm{I}}B_{\mathrm{z}}(t) =
\frac{\mu _{0}}{4\pi }\cdot \gamma _{\mathrm{I}}\gamma
_{\mathrm{S}}\hbar \cdot {\sum_{\mathrm{i}}} \frac{1 -
3z_{\mathrm{i}}^{2}/r_{\mathrm{i}}^{2}}{r_{\mathrm{i}}^{3}}\cdot
(S_{\mathrm{z}}(\mathbf{r}_{\mathrm{i}},t) -
\left<S_{\mathrm{z}}(\mathbf{r}_{\mathrm{i}})\right>),\label{20}
\end{eqnarray}
where $\mu _{0} = 0.4\pi
\,\mathrm{T}^{2}\mathrm{cm}^{3}/\mathrm{J}$,
$\mathbf{r}_{\mathrm{i}}$ is radius-vector of the impurity atom
distance.\par

In this case, once again, we have
\begin{eqnarray}
\left<\Delta \omega (t)\Delta \omega (0)\right> = \left<\Delta
\omega ^{2}\right>\cdot \exp (-t/\tau
_{1,\mathrm{imp}}),\label{21}
\end{eqnarray}
where for non-correlated homogeneous spatial distribution of
impurity atoms
\begin{eqnarray}
1/T_{\mathrm{d}}^{2} \approx \left<\Delta \omega ^{2}\right>
\approx C_{\mathrm{S},\mathrm{imp}}((\mu _{0}/4\pi )\cdot \gamma
_{\mathrm{I}}\gamma _{\mathrm{S}}\hbar )^{2}\cdot \frac{16\pi
}{15a^{3}}\cdot (\left<S_{\mathrm{z}}^{2}\right> -
\left<S_{\mathrm{z}}\right>^{2}),\label{22}
\end{eqnarray}
where
\begin{eqnarray}
\left<S_{\mathrm{z}}^{2}\right> - \left<S_{\mathrm{z}}\right>^{2}
\approx \,\exp (-(\gamma _{\mathrm{S}}\hbar B/kT)),\label{23}
\end{eqnarray}
$C_{\mathrm{S},\mathrm{imp}}\,\mathrm{is}$ concentration of
magnetic atoms in $\mathrm{cm}^{-3}$, $\gamma _{\mathrm{S}} =
176\,\mathrm{rad}\,\mathrm{GHz}/\mathrm{T}$,
$\gamma_{\mathrm{I}}(^{31}\mathrm{P}) =
108\,\mathrm{rad}\,\mathrm{MHz}/\mathrm{T}$, $a$ is minimal
distance of order of lattice constant, for Si $a^{-3} \approx
5.0\cdot 10^{22}\,\mathrm{cm}^{-3}$,
\begin{eqnarray}
\left<\Delta \omega ^{2}\right> \approx 33.4\cdot 10^{-13}\cdot
(C_{\mathrm{S},\mathrm{imp}}a^{3})\cdot \exp (-(\gamma
_{\mathrm{S}}\hbar B/kT)).\label{24}
\end{eqnarray}
It follows that a direct suppression of decoherence due to
interaction with magnetic impurity atoms may be achieved by\par

increase of relation ($B/T$) and\par

decrease of impurity concentration $C_{\mathrm{S},\mathrm{imp}}$.\par

To estimate the allowed concentration of paramagnetic centers in
the silicon substrate $C_{\mathrm{S},\mathrm{imp}}$ we will
consider as previously the case slow frequency modulation and the
decoherence time $T_{\mathrm{dec}} \sim 1\,\mathrm{s}$. For $B/T >
20\,\mathrm{T}/\mathrm{K}$ we obtain
\begin{eqnarray}
1/T_{\mathrm{d}}^{2} \approx \left<\Delta \omega ^{2}\right>
\approx 0.74\cdot 10^{3}\cdot (C_{\mathrm{S},\mathrm{imp}}a^{3}) <
1\,\mathrm{s}^{-2}\label{25}
\end{eqnarray}
or
\begin{eqnarray}
C_{\mathrm{S},\mathrm{imp}} < 1.4\cdot 10^{-3}\cdot a^{-3} \approx 0.7\cdot 10^{20}\,\mathrm{cm}^{-3}.\label{26}
\end{eqnarray}
Thus, the value of allowed concentration of magnetic impurity is
practically unbounded at large enough value $\gamma
_{\mathrm{S}}\hbar B/kT$.\par

Another mechanism of individual qubit state decoherence is
dipole-dipole interaction with nuclear spins $I \neq 0$ of
impurity diamagnetic atoms having concentration
$C_{\mathrm{I},\mathrm{imp}}$. Isotope $^{29}\mathrm{Si}$ with
$\gamma _{\mathrm{I},\mathrm{imp}} = -
53\,\mathrm{rad}\,\mathrm{MHz}/\mathrm{T}$ is one of the such
atoms.\par

In this case correlation function take form
\begin{eqnarray}
\left<\Delta _{\mathrm{S}}\omega (\tau )\Delta \omega
_{\mathrm{S}}(0)\right> = \left<\Delta \omega ^{2}\right>\cdot
\exp (-t/T_{\parallel\mathrm{imp}} ),\ \ \ \ \tau _{\mathrm{C}} =
T_{\parallel,\mathrm{imp}} , \label{27}
\end{eqnarray}
where $T_{\parallel,\mathrm{imp}} \approx 10^{4}\,\mathrm{s} -
$impurity nuclear spin longitudinal relaxation time of isotope
$^{29}\mathrm{P}$ at low temperature \cite{7}.
\begin{eqnarray}
1/T_{\mathrm{d}}^{2} \approx C_{\mathrm{I},\mathrm{imp}}((\mu
_{0}/4\pi )\cdot \gamma _{\mathrm{I}}\gamma
_{\mathrm{I},\mathrm{imp}}\hbar )^{2}\cdot \frac{4\pi
}{15a^{3}}\cdot (1 - \tanh^{2}(|\gamma
_{\mathrm{I},\mathrm{imp}}|\hbar B/kT_{\mathrm{I}})))\label{28}
\end{eqnarray}
For $B = 2\,\mathrm{T}$ and for spin temperature $T_{\mathrm{I}}$ at which there are near-full polarization of impurity nuclear spins
\begin{eqnarray}
|\gamma _{\mathrm{I},\mathrm{imp}}\hbar B/kT_{\mathrm{I}}| > 1\label{29}
\end{eqnarray}
or for $T_{\mathrm{I}} < 0.8\,\mathrm{mK}$, we will obtain that the allowed concentration of
the isotope $^{29}\mathrm{Si}$ in \% is
\begin{eqnarray}
C_{\mathrm{I},\mathrm{imp}}\% < 4.5\cdot 10^{-2}\%\label{30}
\end{eqnarray}
This value may be increased provided a further decrease of spin
temperature $T_{\mathrm{I}}$.\par

For comparison, natural abundance of isotope $^{29}\mathrm{Si}$ in natural
silicon is 4.7\%.\par

Hence the main reasons for the internal decoherence of qubit
states are the modulation of resonance qubit frequency due to
hyperfine interaction with fluctuating electron spin and due to
interaction with randomly distributed impurity diamagnetic atoms
containing nuclear spins.\par

\section{The random phase errors in ensemble quantum register}

The preparation of both the initialized and other basic states of
quantum register is followed by error generation through
interaction of qubits with the environment. Any one-qubit error in
either of the basic state is defined as a superposition basic
states $\left|0\right>$ and $\left|1\right>$.\par

As an example of one-qubit error action we consider the following
unitary (it is not necessary unitary) transformation
\begin{eqnarray}
\mathbf{U}_{\varepsilon } = \frac{1}{\sqrt{1+\varepsilon _{\mathrm{x}}^{2}+\varepsilon _{\mathrm{y}}^{2}+\varepsilon _{\mathrm{z}}^{2}}}\cdot \left[
\begin{tabular}{c c }
 $1 + i\varepsilon _{\mathrm{z}}$ & $i\varepsilon _{\mathrm{x}}+\varepsilon _{\mathrm{y}}$\\
 $i\varepsilon _{\mathrm{x}}-\varepsilon _{\mathrm{y}}$ & $1 - i\varepsilon _{\mathrm{z}}$\\
\end{tabular}
\right],\label{31}
\end{eqnarray}
where $\varepsilon _{\mathrm{x}}$, $\varepsilon _{\mathrm{y}}$,
$\varepsilon _{\mathrm{z}}$ are independent random error function
of time.\par

The action of this transformation on initialized state $\rho (0) =
\left|0\right>\left<0\right| = \left[
\begin{tabular}{c c }
 $1$ & $0$\\
 $0$ & $0$\\
\end{tabular}
\right]$ leads to the modified pure state
\begin{eqnarray}
\rho _{\varepsilon }(0) = \mathbf{U}_{\varepsilon
}\left|0\right>\left<0\right|\mathbf{U}^{-1}_{\varepsilon } =
1/2\cdot \left[
\begin{tabular}{c c }
 $1 + P_{\mathrm{z},\varepsilon }$ & $P_{-,\varepsilon }$\\
 $P_{+,\varepsilon }$ & $1 - P_{\mathrm{z},\varepsilon }$\\
\end{tabular}
\right],\label{32}
\end{eqnarray}
where the components of Bloch vector are
\begin{eqnarray}
P_{\mathrm{z},\varepsilon } = 1 - 2p_{\varepsilon }(0), \ \ \
P_{\pm ,\varepsilon } = 2 \frac{\sqrt{(1+\varepsilon
_{\mathrm{z}}^{2})(\varepsilon _{\mathrm{x}}^{2}+\varepsilon
_{\mathrm{y}}^{2})}}{1+\varepsilon _{\mathrm{x}}^{2}+\varepsilon
_{\mathrm{y}}^{2}+\varepsilon _{\mathrm{z}}^{2}}\cdot \exp (\pm
i\varphi _{\varepsilon }),\label{33}
\end{eqnarray}
\begin{eqnarray}
p_{\varepsilon }(0) = (\varepsilon _{\mathrm{x}}^{2}+\varepsilon
_{\mathrm{y}}^{2})/(1+\varepsilon _{\mathrm{x}}^{2}+\varepsilon
_{\mathrm{y}}^{2}+\varepsilon _{\mathrm{z}}^{2}),\ \ \  \tan
\varphi _{\varepsilon } = (\varepsilon _{\mathrm{x}}\varepsilon
_{\mathrm{y}}+\varepsilon _{\mathrm{z}})/(\varepsilon
_{\mathrm{x}}\varepsilon _{\mathrm{z}}-\varepsilon
_{\mathrm{y}}).\label{34}
\end{eqnarray}
The fidelity of initial state $\left|0\right>$ is determined by
\begin{eqnarray}
F(0) =\,\mathrm{Sp}\rho _{\varepsilon }(0)\rho (0) = 1 - p_{\varepsilon }(0),\label{35}
\end{eqnarray}
where $p_{\varepsilon }(0)$ is the random error probability. The diagonal elements of
the perturbed density matrix $\rho _{\varepsilon }(0)$ depend only on the value of $p_{\varepsilon }(0)$
and have no random phase factor.\par

The random phase factors with arbitrary continuous random phase
$\varphi _{\varepsilon }$ have only non-diagonal elements of
density matrix $\rho _{\varepsilon }(0)$ or the Bloch components
$P_{\pm ,\varepsilon }$.\par

All 2$^{L}$ basic state vectors of quantum register
$\left|n\right>$ are generated by one-qubit unitary
transformations $i\mathrm{NOT}$ which transform the initial basis
states of certain qubits $\left|0\right>$ in state
$\left|1\right>$. The containing error
$i\mathrm{NOT}_{\varepsilon}$ operations leads also to appropriate
random phase factors.\par

The existence of the such uncontrollable phase factors can involve
a problem, which mainly lies, as was noted by S.Kak \cite{11}, in
the inability of the quantum error correction codes (QECC) to
correct this analog-type quantum phase errors as ''they can
potentially correct only bit flips and phase flips and some
combination thereof, which errors represent a small subset of all
the error that can corrupt a quantum state''.\par

We will advance here some arguments in support of the ensemble
approach use in connection with the generation of uncontrollable
phase factors.\par

The producing of random phase factors (known as fasors) may be
considered as a stationary random process realized by random
functions $\varepsilon _{\mathrm{x}}(t)$, $\varepsilon
_{\mathrm{y}}(t)$ and $\varepsilon _{\mathrm{z}}(t)$ possessed by
a statistical ensemble. A time-average of density matrix is
practically unfeasible as it requires prolonged state measurements
of an individual quantum register. But if it is assumed that
random process in ensemble quantum registers is ergodic random
process, it may be suggested to substitute the time average by the
ensemble average. Practically the ensemble averaging will occur by
natural way, when unitary transformations and ensemble quantum
register states measurements are produced at the same time for the
whole ensemble by the same controlling pulse, much as in
bulk-ensemble liquid quantum computer prototype. The
ensemble-averaged reduced density matrix will describe now same
mixed state whose non-diagonal matrix elements will decay due to
decoherence process and have no analog-type random values. Such
states involve no special problems for QECC.\par

The decay of non-diagonal matrix elements comes much rapidly the
more dimension of non-diagonal matrix element blocks. Thus during
the process of preparing of many-qubit quantum register basis
states a rapid decoherence can play the constructive role. The
rapid decoherence may be produced, among other processes, by means
of a sharp decrease, for the short time, of external magnetic
field.\par

As the result mixing degree of reduced density matrix
$\left<p_{\varepsilon }\right>$ may be treat as temperature analog
in a liquid quantum computers. For suited conditions the value
$\left<p_{\varepsilon }\right> \ll 1$. Analogously to the low
temperature many-qubit mixed state the considered state may be
transformed to pseudpure state, but in so doing the entanglement
property of large-scaled state at low temperature will not be
violate \cite{12}. In this case we will have the fully quantum
computing instead of it imitation.\par

I am grateful to K.\ A.\ Valiev for stimulating discussions and an
inestimable support and to V.\ A.\ Kokin for the large technical
help.\par

\section*{Summary}

The main mechanisms of decoherence for low temperature are the
adiabatic processes of random modulation of qubit resonance
frequency without spin flips produced by hyperfine interaction of
nuclear spin with electron spin of the basic atoms and
dipole-dipole interaction with nuclear spins of impurity
diamagnetic atoms. It is estimated the allowed concentration of
nuclear spin containing isotopes $^{29}\mathrm{P}$. It was shown
that the random phase error of quantum register state in the
ensemble approach are averaged and the pure basic state transforms
to the mixed state, analogical to mixed state at nonzero
temperatures.\par
\par


\begin{thebibliography}{99}

\bibitem{1} DiVincenzo,\ D.\ P., The Physical Implementation of Quantum Computation, {\it Fortschr.\ Phys.}, 2000, vol.\ 48, no\ 9-11, pp.\ 771--783.\par
\bibitem{2} Kane,\ B.\ E., A Silicon-Based Nuclear Spin Quantum Computer, {\it Nature} (London), 1998, vol.\ 393, pp.\ 133--137.\par
\bibitem{3} Kane,\ B.\ E., Silicon-based Quantum Computation, {\it Fortschr.\ Phys.}, 2000, vol.\ 48, no\ 9-11, pp.\ 1023--1041.\par
\bibitem{4} Valiev,\ K.\ A., Kokin, A.\ A., Solid-State NMR Quantum Computer with Individual Access to Qubits and Some Their Ensemble Developments, E-print LANL: quant-ph/9909008, 1999.\par
\bibitem{5} Valiev,\ K.\ A., Kokin, A.\ A., Quantum Computers: Reliance and Reality., Moscow-Izhevsk: R\&C Dynamics, 2001 (in Russian).\par
\bibitem{6} Kokin,\ A.\ A., Valiev, K.\ A., Problems in Realization of Large-Scale Ensemble Silicon-Based NMR Quantum Computers, E-print LANL: arXiv:quant-ph/0201083, 2002.;\par
\bibitem{7} Feher,\ G, Gere,\ E.\ A., Electron Spin Resonance Experiments on Donors in Silicon.\ II.\ Electron Spin Relaxation Effects, {\it Phys.\ Rev.}, 1959, vol.\ 114, no.\ 5, pp.\ 1245--1256.\par
\bibitem{8} Abragam,\ A., Goldman,\ M., Nuclear Magnetism: Order \& Disorder. Oxford: Clarend.Press, 1982.\par
\bibitem{9} Wellard,\ C.\ J., Hollenberg,\  L.\ C.\ L., Stochastic Noise as a Source of Decoherence in a Solid State Quantum Computer, E-print LANL: arXiv:quant-ph/0104055, 2001.\par
\bibitem{10} Alicki,\ R. Decoherence in Quantum Open Systems Revisited, E-print LANL: arXiv:quant-ph/0205173, 2002.\par
\bibitem{11} Kak,\ S., General Qubit Errors Cannot be Corrected, E-print LANL: arXiv:quant-ph/0206144, 2002.\par
\bibitem{12} D\"{u}r,\ W., Cirac,\ C.\ I., Tarrach,\ R. Separability and Distrillability of Multiparticle Quantum Systems, {\it Phys.\ Rev.\ Lett.}, 1999, vol.\ 83, no.\ 17, pp.\ 3562--3565.\par

\end{thebibliography}
\end{document}